\documentclass[superscriptaddress, twocolumn]{revtex4}

\usepackage[pdftex]{graphicx}

\begin{document}

\title{High-speed measurement of rotational anisotropy nonlinear optical harmonic generation using position sensitive detection}

\author{J. W. Harter$^{\dagger}$}
\affiliation{Department of Physics, California Institute of Technology, Pasadena, CA 91125, USA}
\affiliation{Institute for Quantum Information and Matter, California Institute of Technology, Pasadena, CA 91125, USA}

\author{L. Niu$^{\dagger}$}
\affiliation{Department of Physics, California Institute of Technology, Pasadena, CA 91125, USA}
\affiliation{Institute for Quantum Information and Matter, California Institute of Technology, Pasadena, CA 91125, USA}

\author{A. J. Woss}
\affiliation{Department of Applied Mathematics and Theoretical Physics, University of Cambridge, Cambridge, CB3 0WA, UK}

\author{D. Hsieh}
\email[Author to whom correspondence should be addressed: ]{dhsieh@caltech.edu}
\affiliation{Department of Physics, California Institute of Technology, Pasadena, CA 91125, USA}
\affiliation{Institute for Quantum Information and Matter, California Institute of Technology, Pasadena, CA 91125, USA}

\date{\today}

\begin{abstract}
We present a method of performing high-speed rotational anisotropy nonlinear optical harmonic generation experiments at rotational frequencies of several hertz by projecting the harmonic light reflected at different angles from a sample onto a stationary position sensitive detector. The high rotational speed of the technique, $10^3$ to $10^4$ times larger than existing methods, permits precise measurements of the crystallographic and electronic symmetries of samples by averaging over low frequency laser power, beam pointing, and pulse width fluctuations. We demonstrate the sensitivity of our technique by resolving the bulk four-fold rotational symmetry of GaAs about its [001] axis using second harmonic generation.
\end{abstract}

\maketitle

\begin{figure*}[t]
\includegraphics[width=\linewidth]{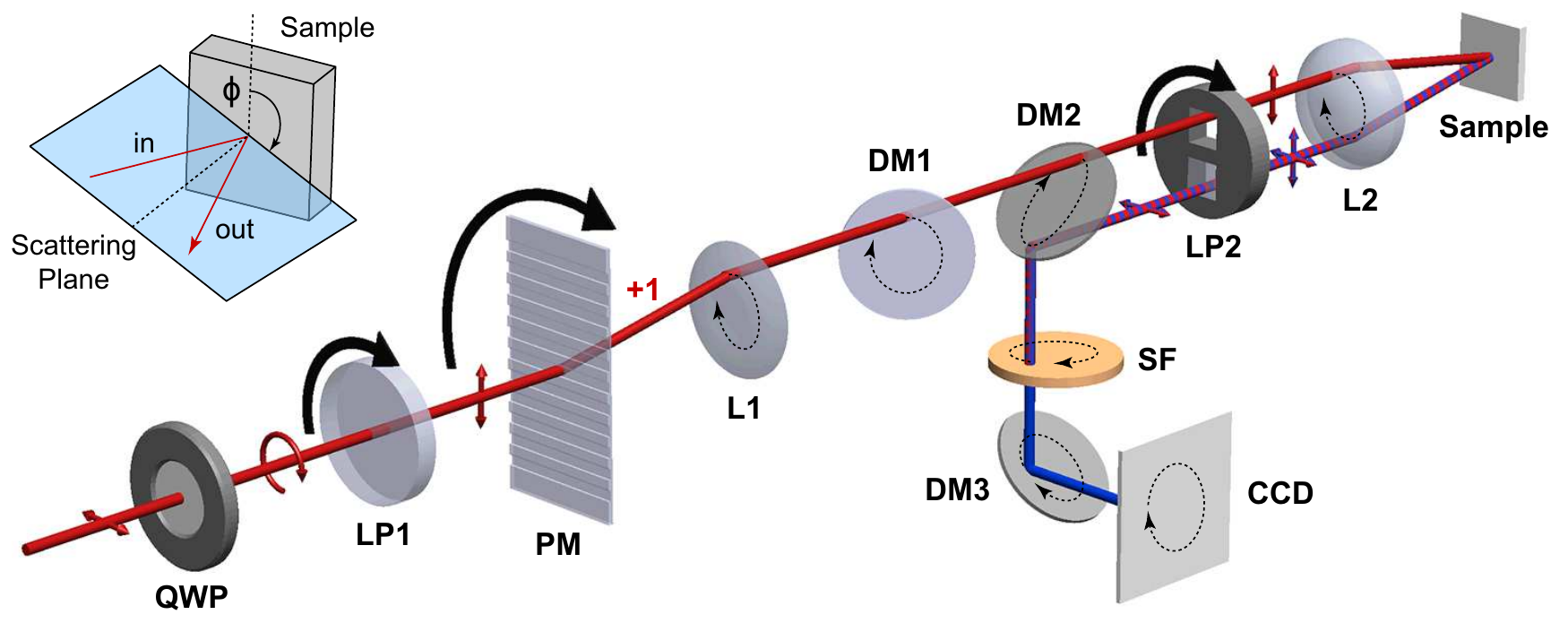}
\caption{\label{fig:Fig1} Schematic layout of the high-speed RA-NHG setup. Laser light is incident from the left and passes through a series of optics labeled as follows: quarter-wave plate (QWP), linear polarizers (LP1,2), binary phase mask (PM), collimating lens (L1), dichroic mirrors (DM1-3), achromatic objective lens (L2), spectral filter (SF), and CCD camera (CCD). LP2 is mounted on one side of a rotating wheel while the other side has a through-hole. Only the +1 diffracted order from the PM used in our experiment is shown. The polarization of the fundamental (red) and higher harmonic (blue) light is illustrated at various points along the beam path. The PM and two LPs are mounted on a motorized rotation stage (not shown) and are rotated co-axially (solid curved arrows). The dashed arrows show the path traced by the beam on L1, L2, DM1-3, SF, and CCD during the rotation. Top left inset shows the simplified geometry of a reflection-based RA-NHG measurement.}
\end{figure*}

Rotational anisotropy nonlinear harmonic generation (RA-NHG) is an all-optical technique for determining the crystallographic and electronic symmetries of solids \cite{Boyd}. It has been successfully applied to study a variety of phenomena, including lattice reconstruction on semiconductor thin film surfaces \cite{Yamada,Tom1,Heinz,Lee} and metal/electrolyte interfaces \cite{Shannon,Shen_review}, magnetic ordering on metal surfaces and buried multilayer interfaces \cite{Pan,Reif,Dahn,Gridnev,Kirilyuk_Review,Nyvlt}, magnetic ordering in bulk semiconductors \cite{Lafrentz} and multiferroics \cite{Fiebig_Cr2O3,Fiebig_manganite,Fiebig_Review,Orenstein}, and molecular self-assembly on dielectric substrates \cite{Bilderling}. In the RA-NHG technique, a laser beam is focused onto a sample and the intensity of light reflected or transmitted at a higher harmonic of the incident frequency is measured as a function of the angle $\phi$ between the scattering plane and some crystalline axis parallel to the surface of the sample while a fixed angle of incidence is maintained [inset Fig.~\ref{fig:Fig1}]. By selectively measuring different combinations of incoming and outgoing light polarizations, typically linearly polarized either in ($P$) or out of ($S$) the scattering plane, the independent non-zero elements of the nonlinear optical susceptibility tensor of the sample can be deduced. By Neumann's principle, this serves to identify the crystallographic and electronic point group symmetries of the sample under study \cite{Boyd}.

\begin{figure}[t]
\includegraphics[width=\linewidth]{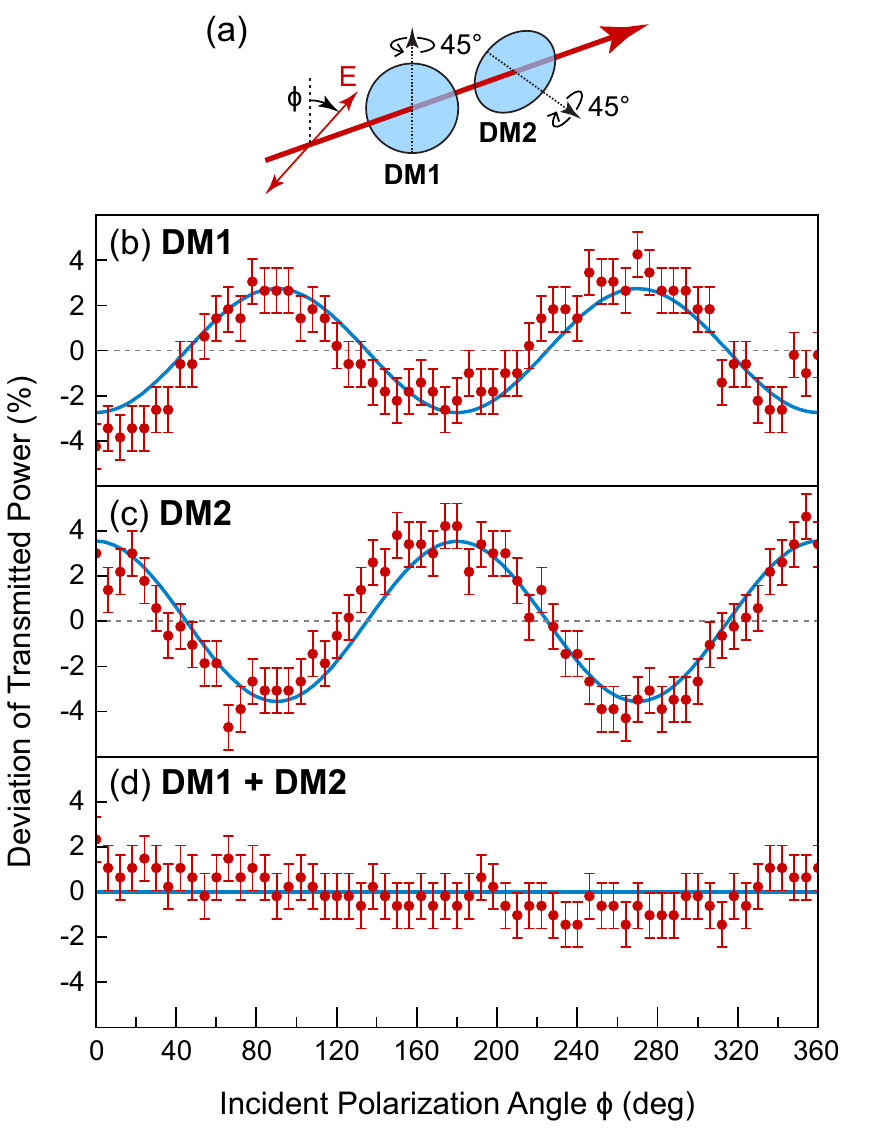}
\caption{\label{fig:Fig2} Demonstration of polarization compensation by a dichroic mirror pair. (a) Polarization-dependent transmittance curves of DM1 and DM2 were measured using a photodiode positioned after DM2 with both PM and L1 removed. Data were acquired by stepping the polarization angle $\phi$ in $6^{\circ}$ increments and averaging for $\sim 10$ sec per data point. The transmitted intensity of 800 nm light through (b) only DM1, (c) only DM2, and (d) both DM1 and DM2 are displayed as a function of $\phi$. Also shown are fits to $\cos\phi$ for the individual mirrors and a constant for the mirror pair. Fluctuations of the data about the fitting functions are due primarily to low frequency laser power fluctuations during acquisition. The error bars show the RMS deviation of the data from the fits.}
\end{figure}

RA-NHG experiments are typically carried out by rotating the sample while keeping the scattering plane fixed. An alternative approach in which the scattering plane is rotated while the sample is kept fixed has also recently been demonstrated, which is advantageous when studying small samples and when low temperatures or externally applied fields (e.g. magnetic, electric, strain) are desired \cite{Torchinsky_RSI,Torchinsky_PRL}. In both methods, $\phi$ is stepped and held at discrete values for a finite wait time while the high harmonic intensity is recorded. A major drawback of this method of data acquisition is the long time required (of order $10^3$ sec) to complete a full 360$^{\circ}$ sweep of $\phi$. This introduces a large amount of low frequency noise in the data from temporal fluctuations in laser power, beam pointing, and pulse width, limiting the ultimate precision with which point group symmetries can be determined. It also precludes accurately measuring temporal changes in the RA-NHG response of a sample that occur on the timescale of a $\phi$ rotation period, which may arise from processes such as charge migration or impurity adsorption on the sample surface under vacuum conditions \cite{Tom2,Hsieh_SHG}. In this Letter, we demonstrate a novel high-speed RA-NHG setup using a position sensitive detector that rotates $\phi$ at a frequency $10^3$ to $10^4$ times higher than conventional methods \cite{Torchinsky_RSI}, which can in principle be increased to even higher frequencies.

A schematic layout of the high-speed RA-NHG setup is shown in Fig.~\ref{fig:Fig1}. Unless otherwise stated, all optical components are standard and were obtained from Thorlabs, Inc. The setup operates in the following way: A circularly polarized laser beam, produced from a linearly polarized beam using a quarter-wave plate (QWP), first passes through a linear polarizer (LP1) and a fused silica binary phase mask (PM; Tessera Technologies, Inc.) that are mechanically co-rotating about the beam axis. The polarizer produces a beam of constant intensity with a rotating linear polarization that is maintained either parallel or perpendicular to the PM groove direction depending on whether an $S_{in}$ or $P_{in}$ polarization geometry at the sample is required. The phase mask splits the beam into many diffracted orders, each retaining the polarization set by LP1 and sweeping out a cone as the PM rotates. For our measurements, only the $+1$ order is utilized while all other orders are blocked. Next, a collimating lens (L1) and an achromatic objective lens (L2) form a telescope that focuses the beam onto the surface of the sample, thus creating a rotating scattering plane. The fundamental and higher harmonic beams reflected from the sample return via the diametrically opposite side of L2 upon which they are re-collimated. They then pass through a linear polarizer (LP2) that is co-rotating with both LP1 and PM, which is used to select either the $S_{out}$ or $P_{out}$ polarization component of the reflected beam. The desired harmonic is finally isolated using tilted dichroic mirrors (DM2,3; Semrock, Inc.\ BrightLine series) and spectral filters (SF) that direct it onto a cooled electron-multiplying CCD camera (CCD; Andor Technology Ltd.). As LP1, PM, and LP2 rotate, the reflected beam sweeps out a circle on the two-dimensional active area of the CCD, thereby recording the angular $\phi$-dependence of the NHG signal as a position dependent intensity on the CCD. To eliminate any possibility of mechanical phase slip between LP1, PM, and LP2 during rotation, they are coupled to a common axle driven by a single 12 V DC electric motor with a pulse-width modulating power controller. The need to rotate only three lightweight optical elements allows for rotation at high speeds ($\sim 4$ Hz or more), which is greater than previous setups that rotate the detector \cite{Torchinsky_RSI,Torchinsky_PRL} by a factor of $10^3$ to $10^4$ and much faster than the characteristic timescale for laser fluctuations and drift. Moreover, because the value of $\phi$ read from the CCD image is rigidly tied to the angle of the PM, the experiment is insensitive to random phase slips of the motor.

The key component of our high-speed RA-NHG setup, enabling the use of a stationary detector, is the triple dichroic mirror periscope (DM1-3). This consists of three identical long-pass dichroic mirrors with a cut-on edge wavelength between that of the fundamental and desired harmonic, each mounted at 45$^{\circ}$ with respect to the direction of beam propagation, as shown in Fig.~\ref{fig:Fig1}. To illustrate the function of this periscope, we first study its effect on the fundamental beam by measuring the intensity transmitted through DM1, DM2, and both mirrors as a function of the incident linear polarization angle $\phi$ [Fig.~\ref{fig:Fig2}(a)]. As $\phi$ changes, the polarization incident on a dichroic mirror that is rotated by 45$^{\circ}$ about the vertical axis varies continuously between $S$ and $P$ with a $180^{\circ}$ periodicity. The transmittance through DM1 therefore varies sinusoidally in accordance with the Fresnel equations, as plotted in Fig.~\ref{fig:Fig2}(b). In order to compensate for this unwanted effect, the beam is sent through a second dichroic mirror (DM2) that is tilted by 45$^{\circ}$ about the horizontal axis, which has a transmittance curve that is $180^{\circ}$ out of phase with the first mirror [Fig.~\ref{fig:Fig2}(c)]. The combined effect of DM1 and DM2 is to completely eliminate any net polarization dependence of the transmittance [Fig.~\ref{fig:Fig2}(d)] and thereby maintain a constant $\phi$-independent incident intensity on the sample. The combination of DM2 and DM3 acts in a similar way in reflection geometry to eliminate any net polarization dependence of the outgoing high harmonic beam. We note that the acquisition time for each curve shown in Fig.~\ref{fig:Fig2} is approximately 10 minutes. On this timescale, characteristic low frequency laser fluctuations are clearly observed through the deviations of the data away from the fitted curves and the mismatch of the $\phi = 0^{\circ}$ and $360^{\circ}$ data points. It is precisely these types of fluctuations that are averaged out by the high-speed RA-NHG setup.

\begin{figure*}[t]
\includegraphics[width=\linewidth]{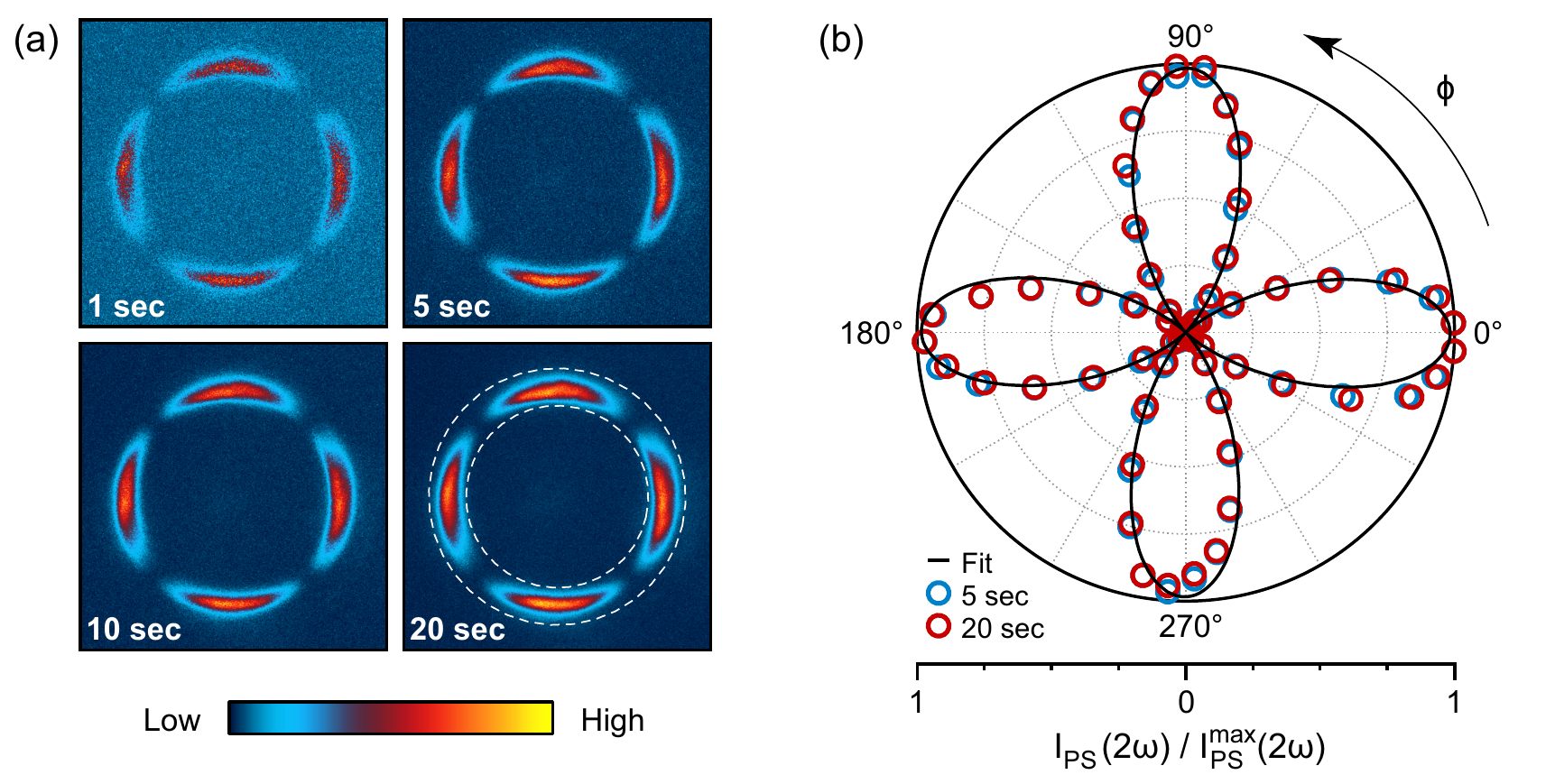}
\caption{\label{fig:Fig3} RA-SHG data collected from (001) GaAs using the high speed RA-NHG setup, obtained with 800 nm incident light under a $P_{in}$-$S_{out}$ polarization geometry and a fluence of $\sim 5$ mJ/cm$^2$. (a) CCD images with 1, 5, 10, and 20 sec exposure times. The GaAs crystal was aligned such that $\phi = 0^{\circ}$ corresponds to the [100] axis. (b) Polar plot of the RA-SHG intensity obtained by performing a radial integration of the 5 sec and 20 sec data in panel (a) over the interval bounded by the concentric white circles, with an azimuthal bin size of $\Delta\phi = 6^{\circ}$. A fit to the expected functional form derived in Ref.~\citenum{Yamada} is overlaid. We note that because the camera exposure times are not locked to integer multiples of the $\phi$ revolution period, there will be a partial range of angles that acquire one extra revolution's worth of signal. This effect can be corrected by reducing the overshot data points by a factor of 1/$N$ where $N$ = (exposure time $\times$ rotation frequency) is the total number of revolutions acquired. Such corrections are only important for short exposure times and were applied to the 5 sec data between 98$^{\circ}$ and 134$^{\circ}$.}
\end{figure*}

To characterize the performance of our RA-NHG setup, we measured the rotational anisotropy of second harmonic generation (SHG) from the (001) crystal surface of GaAs. For the incident light source we used a regeneratively amplified Ti:sapphire laser system (KMLabs Wyvern 1000) that produces 60 fs optical pulses (broadened to $\sim 150$ fs at the sample by dispersion) with 800 nm center wavelength and operates at a 10 kHz repetition rate. The dichroic mirrors were selected to have a cut-on wavelength of 635 nm. At a rotation frequency of $(1/360^{\circ})d\phi/dt \approx$ 4 Hz, the angular spacing between successive pulses is $\sim 0.14^{\circ}$, which is much smaller than the angular subtense of each beam spot on the CCD ($\sim 9^{\circ}$) and therefore allows the incident beam to be treated as quasi-CW. Raw CCD images of the RA-SHG data (400 nm wavelength) from (001) GaAs obtained in a $P_{in}$-$S_{out}$ geometry are shown in Fig.~\ref{fig:Fig3}(a) for a series of short continuous camera exposure times. A ring of intensity with alternating nodes and antinodes separated by 45$^{\circ}$ is already clearly visible after 1 sec, which demonstrates a rotationally anisotropic SHG response of the sample.

The CCD images can be cast into a form more conducive to quantitative analysis by performing a radial integration (after background subtraction) with respect to an origin centered on the ring of intensity and binning the data into fixed $\phi$ azimuthal intervals. Figure~\ref{fig:Fig3}(b) shows the integrated intensity of both the 5 sec and 20 sec images as a function of $\phi$ on a polar plot, which exhibit four lobes of equal amplitude consistent with previously reported RA-SHG measurements on the (001) GaAs surface \cite{Yamada}. Quantitatively, the data are also well described by the expression $|\textrm{A}\cos(2\phi)|^2$ expected under $P_{in}$-$S_{out}$ geometry, where $\textrm{A}$ is a linear combination of bulk and surface electric dipole induced SHG susceptibility tensor elements derived from the $C_{2v}$ symmetry group of GaAs \cite{Yamada}. A closer examination of the data reveals a high degree of overlap between data points taken at different exposure times, which demonstrates the high level of precision achievable using our technique and shows that the minor deviations of the data away from the ideal $|\textrm{A}\cos(2\phi)|^2$ form are due to systematic sources of error such as misalignment and sample surface quality rather than statistical sources of error such as laser fluctuations.

In summary, we have demonstrated a novel technique based on position sensitive detection to perform RA-NHG experiments at rotational frequencies $10^3$ to $10^4$ times higher than existing methods. The frequency can in principle be pushed much higher using a more powerful motor \cite{Morris}, which would further suppress low frequency noise arising from laser fluctuations and allow increasingly sensitive measurements of point group symmetries. This technique can be applied in both reflection and transmission geometries and can be extended over a wide photon wavelength range, limited primarily by the spectral range of the CCD. It is also amenable to ultrafast time-resolved pump-probe RA-NHG measurements through the use of short laser pulses and the introduction of a pump beam, which can be used for example to separate electronic relaxation pathways \cite{Hsieh_Mahmood} or to study photo-induced symmetry breaking phase transitions \cite{Cavalleri}. The time resolution of such experiments can be improved by replacing the transmissive objective L2 with a reflective Cassegrain objective to reduce dispersive pulse broadening, and sensitivity to small pump-induced changes in NHG intensity can be enhanced with a lock-in camera.

$^{\dagger}$ These authors contributed equally to this work.

\section*{Funding Information}

This work is supported by the U. S. Department of Energy under award number DE-SC0010533. Instrumentation for the NHG-RA setup was partially supported by a U. S. Army Research Office DURIP award under grant number W911NF-13-1-0293. D.H. acknowledges funding provided by the Alfred P. Sloan Foundation (FG-BR2014-027) and the Institute for Quantum Information and Matter, an NSF Physics Frontiers Center (PHY-1125565) with support of the Gordon and Betty Moore Foundation through Grant GBMF1250. A.J.W. acknowledges support from the Caltech-University of Cambridge SURF exchange.

\section*{Acknowledgments}

The authors thank Darius Torchinsky for suggesting the idea of a triple dichroic mirror periscope.


\begin{thebibliography}{99}

\bibitem{Boyd} R. W. Boyd, \textit{Nonlinear optics} (Academic Press, 1991).

\bibitem{Yamada} C. Yamada and T. Kimura, Phys. Rev. Lett. \textbf{70}, 2344 (1993).

\bibitem{Tom1} H. W. K. Tom, T. F. Heinz, and Y. R. Shen, Phys. Rev. Lett. \textbf{51}, 1983 (1983).

\bibitem{Heinz} T. F. Heinz, M. M. T. Loy, and W. A. Thompson, Phys. Rev. Lett. \textbf{54}, 63 (1985).

\bibitem{Lee} Y. S. Lee, M. H. Anderson, and M. C. Downer, Opt. Lett. \textbf{22}, 973 (1997).

\bibitem{Shannon} V. L. Shannon, D. A. Koos, and G. L. Richmond, Appl. Opt. \textbf{26}, 3579 (1987); J. Phys. Chem. \textbf{91}, 5548 (1987).

\bibitem{Shen_review} Y. R. Shen, Ann. Rev. Phys. Chem. \textbf{40}, 327 (1989).

\bibitem{Pan} R.-P. Pan, H. D. Wei, and Y. R. Shen, Phys. Rev. B \textbf{39}, 1229 (1989).

\bibitem{Reif} J. Reif, J. C. Zink, C.-M. Schneider, and J. Kirschner, Phys. Rev. Lett. \textbf{67}, 2878 (1991).

\bibitem{Dahn} A. Dahn, W. Hubner, and K. H. Bennemann, Phys. Rev. Lett. \textbf{77}, 3929 (1996).

\bibitem{Gridnev} V. N. Gridnev, V. V. Pavlov, R. V. Pisarev, A. Kirilyuk, and T. Rasing, Phys. Rev. B \textbf{63}, 184407 (2001).

\bibitem{Kirilyuk_Review} A. Kirilyuk and T. Rasing, J. Opt. Soc. Am. B \textbf{22}, 148 (2005).

\bibitem{Nyvlt} M. Nyvlt, F. Bisio, and J. Kirschner, Phys. Rev. B \textbf{77}, 014435 (2008).

\bibitem{Lafrentz} M. Lafrentz, D. Brunne, B. Kaminski, V. V. Pavlov, R. V. Pisarev, A. B. Henriques, D. R. Yakovlev, G. Springholz, G. Bauer, and M. Bayer, Phys. Rev. B \textbf{85}, 035206 (2012).

\bibitem{Fiebig_Cr2O3} M. Fiebig, D. Fr\"{o}hlich, B. B. Krichevtsov, and R. V. Pisarev, Phys. Rev. Lett. \textbf{73}, 2127 (1994).

\bibitem{Fiebig_manganite} M. Fiebig, D. Fr\"{o}hlich, K. Kohn, St. Leute, Th. Lottermoser, V. V. Pavlov, and R. V. Pisarev, Phys. Rev. Lett. \textbf{84}, 5620 (2000).

\bibitem{Fiebig_Review} M. Fiebig, V. V. Pavlov, and R. V. Pisarev, J. Opt. Soc. Am. B \textbf{22}, 96 (2005).

\bibitem{Orenstein}  A. Kumar, R. C. Rai, N. J. Podraza, S. Denev, M. Ramirez, Y.-H. Chu, L. W. Martin, J. Ihlefeld, T. Heeg, J. Schubert, D. G. Schlom, J. Orenstein, R. Ramesh, R. W. Collins, J. L. Musfeldt, and V. Gopalan, Appl. Phys. Lett. \textbf{92}, 121915 (2008).

\bibitem{Bilderling} C. von Bilderling, M. Tagliazucchi, E. J. Calvo, and A. V. Bragas, Opt. Express \textbf{17}, 10642 (2009).

\bibitem{Torchinsky_RSI} D. H. Torchinsky, H. Chu, T. Qi, G. Cao, and D. Hsieh, Rev. Sci. Instrum. \textbf{85}, 083102 (2014).

\bibitem{Torchinsky_PRL} D. H. Torchinsky, H. Chu, L. Zhao, N. B. Perkins, Y. Sizyuk, T. Qi, G. Cao, and D. Hsieh, Phys. Rev. Lett. \textbf{114}, 096404 (2015).

\bibitem{Hsieh_SHG} D. Hsieh, J. W. McIver, D. H. Torchinsky, D. R. Gardner, Y. S. Lee, and N. Gedik, Phys. Rev. Lett. \textbf{106}, 057401 (2011).

\bibitem{Tom2} H. W. K. Tom, C. M. Mate, X. D. Zhu, J. E. Crowell, T. F. Heinz, G. A. Somorjai, and Y. R. Shen, Phys. Rev. Lett. \textbf{52}, 348 (1984).

\bibitem{Morris} C. M. Morris, R. Vald\'{e}s Aguilar, A. V. Stier, and N. P. Armitage, Opt. Express \textbf{20}, 12303 (2012).

\bibitem{Hsieh_Mahmood} D. Hsieh, F. Mahmood, D. H. Torchinsky, G. Cao, and N. Gedik, Phys. Rev. B \textbf{86}, 035128 (2000).

\bibitem{Cavalleri} A. Cavalleri, Th. Dekorsy, H. H. W. Chong, J. C. Kieffer, and R. W. Schoenlein, Phys. Rev. B \textbf{70}, 161102(R) (2004).

\end{thebibliography}
\end{document}